\documentclass[final,5p,times,twocolumn]{elsarticle}
\usepackage{amsmath}
\usepackage[colorlinks,urlcolor=blue,linkcolor=cyan,anchorcolor=green,citecolor=blue,bookmarks]{hyperref}
\begin{document}

\begin{frontmatter}

\title{BQA: A High-performance Quantum Circuits Scheduling Strategy Based on Heuristic Search}

\author{Xin-miao Chen}
\ead{cxmghost@163.com}

\author{Shi Wang}
\ead{wangshiphys@foxmail.com}

\author{Yong-jin Ye}
\ead{yeyongjingcdx@126.com}

\author{Bo Jiang}
\ead{b26jiang@126.com}

\author{Yong-zheng Wu \corref{cor}}
\ead{782613169@qq.com}

\cortext[cor]{Corresponding author}

\address{The 32nd Research Institute of China Electronics Technology Group Corporation, 201808, Shanghai, China}

% \affiliation{
%     organization={The 32nd Research Institute of China Electronics Technology Group Corporation},
%     postcode={201808},
%     city={Shanghai},
%     country={China}
% }

\begin{abstract}
    Currently, quantum computing is developing at a high speed because its high parallelism and high computing power bring new solutions to many fields. However, due to chip process technology, it is difficult to achieve full coupling of all qubits on a quantum chip, so when compiling a quantum circuit onto a physical chip, it is necessary to ensure that the two-qubit gate acts on a pair of coupled qubits by inserting swap gates. It will cause great additional cost when a large number of swap gates are inserted, leading to the execution time of quantum circuits longer. In this paper, we designed a way based on the business to insert swap gates BQA(Busy Qubits Avoid). We exploit the imbalance of the number of gates on qubits, trying to hide the overhead of swap gates. At the same time, we also expect swap gates to make as little negative impact on subsequent two-qubit gates as possible. We have designed a heuristic function that can take into account both of these points. Compared with qiskit, the execution time of the circuit optimized by our proposed method is only 0.5 times that of the qiskit compiled circuit. And when the number of two-qubit gates is large, it will achieve higher level than general conditions. This implies higher execution efficiency and lower decoherence error rate.
\end{abstract}


\end{frontmatter}

\section{\label{sec:Introduction}INTRODUCTION}
Since the concept of quantum computing was proposed \cite{BF02650179}, researchers around the world have been committed to promoting the practical application of quantum computers, especially in recent years, great progress in the field of quantum computing has been made \cite{spe3039,Fedorov2022}. By utilizing the fundamental principles of quantum mechanics, such as superposition, interference, and entanglement, quantum computers have inherent parallelism and may outperform classical computers in many computational problems, such as machine learning \cite{Nature549}, chemical simulation \cite{acs.chemrev.8b00803}, code breaking \cite{clarke2020quantum}, financial analysis \cite{ORUS2019100028}, cloud service \cite{rahaman2016overview} and \textit{etc}. However, there is still a huge gap between quantum circuits and actually executable quantum instructions, because quantum circuits are only logical representations of quantum algorithms and do not take into account the various constraints imposed by quantum chips, such as the basic gate-set supported by a quantum chip, the connectivity bewteen individual qubits, the coherence time of each qubit, \textit{etc}.To solve these problems, many organizations have conducted research on quantum compilation, such as IBM’s qiskit \cite{aleksandrowicz2019qiskit}, Google’s cirq \cite{hancockcirq}, Intel’s qHiPSTER \cite{smelyanskiy2016qhipster} and so on. Firstly, quantum compiler convert quantum gates in a quantum algorithm to the basic gate-set supported by the target quantum chip. Then quantum compiler is employed to map logical-qubits to physical-qubits. During the mapping process, there is a serious problem with inserting SWAP gates. Under existing technical conditions, it is difficult to achieve full direct coupling of all qubits on a quantum chip, whereas two-qubits gate in the logical quantum circuit may act on any two qubits \cite{10.1145/3168822, veldhorst2015two}. A simple and straightforward solution is to swap the quantum states of the two qubits involved in the two-qubits gate operation to two qubits with direct coupling by inserting a series of SWAP gates \cite{7422167}, then perform the operation on these two qubits, after the operation, reverse the swap path. However, in most quantum computers, there is no direct implementation of the SWAP gate \cite{doi:10.1142/S0219749912500347}, and it is usually necessary to further decompose the SWAP gate into a combination of basic gate-set. Inserting SWAP gates not only affects the efficiency of program execution, but also the reliability of the program result. Excessively long program execution time will introduce decoherence error, which in turn will lead to unreliable result. To ensure the performance and reliability of quantum circuit, it is necessary to optimize the way to insert SWAP gate.

Therefore, the method of inserting SWAP gates has become one of the focuses of quantum compiler research. Lei Liu and Xinglei Dou availably decreased the number of SWAP gates by inserting SWAP gates between multiple programs \cite{liu2021qucloud}. Zulehner et al., used A-star algorithm optimizing SWAP gates inserting for concurrent CNOT gates \cite{Wille2019MappingQC}. Chi Zhang et al. proposed a time-optimal SWAP insertion scheme based on heuristic search \cite{zhang2021time}. Wille work at global gates mapping optimization with SAT solver\cite{Wille2019MappingQC}. The above studies successfully enhanced both performance and reliability by decreasing the number of SWAP gates and the execution time of quantum circuit. Compared with the above methods, we not only consider the number of gates in the entire quantum circuit, but also pay more attention to the busyness of each qubit, that is, the number of quantum gates on each qubit. In a quantum circuit, the number of quantum gates on each qubit is different, and even for some algorithms, the number of quantum gates on individual busy qubits far exceeds that of other qubits.  studies, we paid more attention to business of qubits.In a quantum circuit, qubits’ business is imbalance, for some algorithms, the distance of business of different qubits may be strongly huge. In a pipeline, the length of the pipeline depends on the longest state. Similar to pipeline, the execution time of a quantum circuit depends on the busiest qubit. Therefore, it is a significant way to optimize quantum circuits by making the busiest qubit’s execution time as short as possible. However, with the SWAP gates inserted, business of qubits can not remain the same.

In this paper, we designed a strategy based on the business qubits to insert swap gates called \emph{Busy-Qubits-Avoid} (BQA) strategy. The rest of the paper is organized as follows. First, we present the background and motivation of the swap inserting in Section \ref{sec:Background}. Then, we introduce the architecture of BQA in Section \ref{sec:ArchitectureOfBQA}. In Section \ref{sec:HeuristicSearch}, we describe the heuristic search method in BQA. Afterward, Section \ref{sec:Evaluation} mainly include the experiment and evaluation of our method. Finally, we summarize our findings in Section \ref{sec:Conclusion}.

\section{\label{sec:Background}BACKGROUND}
\subsection{Quantum Gate}
The base of quantum computing is quantum state, every qubit can be $|0\rangle, |1\rangle$ or a superposition state $|\psi\rangle = \alpha |0\rangle + \beta |1\rangle$ We can perform computation by manipulating the state of qubits with quantum gates. In a quantum computer, each operation is achieved by quantum gates. In general, quantum gates can be divided into three kinds: single-qubit gate, two-qubit gate and muti-qubit gate. For single-qubit gate, each gate operation is considered as a rotation of the unit vector on the Bloch-sphere that represent the quantum state. Thus, any single-qubit gate can be decomposed into a combination of $R_x$, $R_y$, and $R_z$ gates that correpsond to rotating around $x$-axis, $y$-axis, and $z$-axis, respectively. As for two-qubit gates, CNOT gate is of particular interest from a theoretical perspective. Any two-qubit gate can be decomposed into a series of CNOT operations by using the ``Krauss-Cirac decomposition'' \cite{NielsenChuang2010,PhysRevA.52.3457,Williams2011}. In fact, the ability to rotate about arbitrary axes on the Bloch-sphere and the operation to entangle any two-qubit are sufficient to implement an arbitrary quanutm logic, i.e., to approximated any multi-qubit gate \cite{NielsenChuang2010,PhysRevA.52.3457}.

\subsection{Qubit Allocation}
When we design the quantum circuit to implement a quantum algorithm, we can apply any gates on any qubits because we don't need to consider the physical topology of quantum chips. However, these quantum circuits generally cannot be executed directly on qunatum chips. The topology of IBM's \emph{almaden} is abstracted into a graph. In this graph, each vertex represents a qubit and each edge between two vertexes means these two qubits are directly coupled.

\subsection{\label{subsec:ChallengesOfSWAPInserting}Challenges of SWAP Inserting}
\begin{figure*}[htp]
    \centering
    \includegraphics[width=0.90\textwidth]{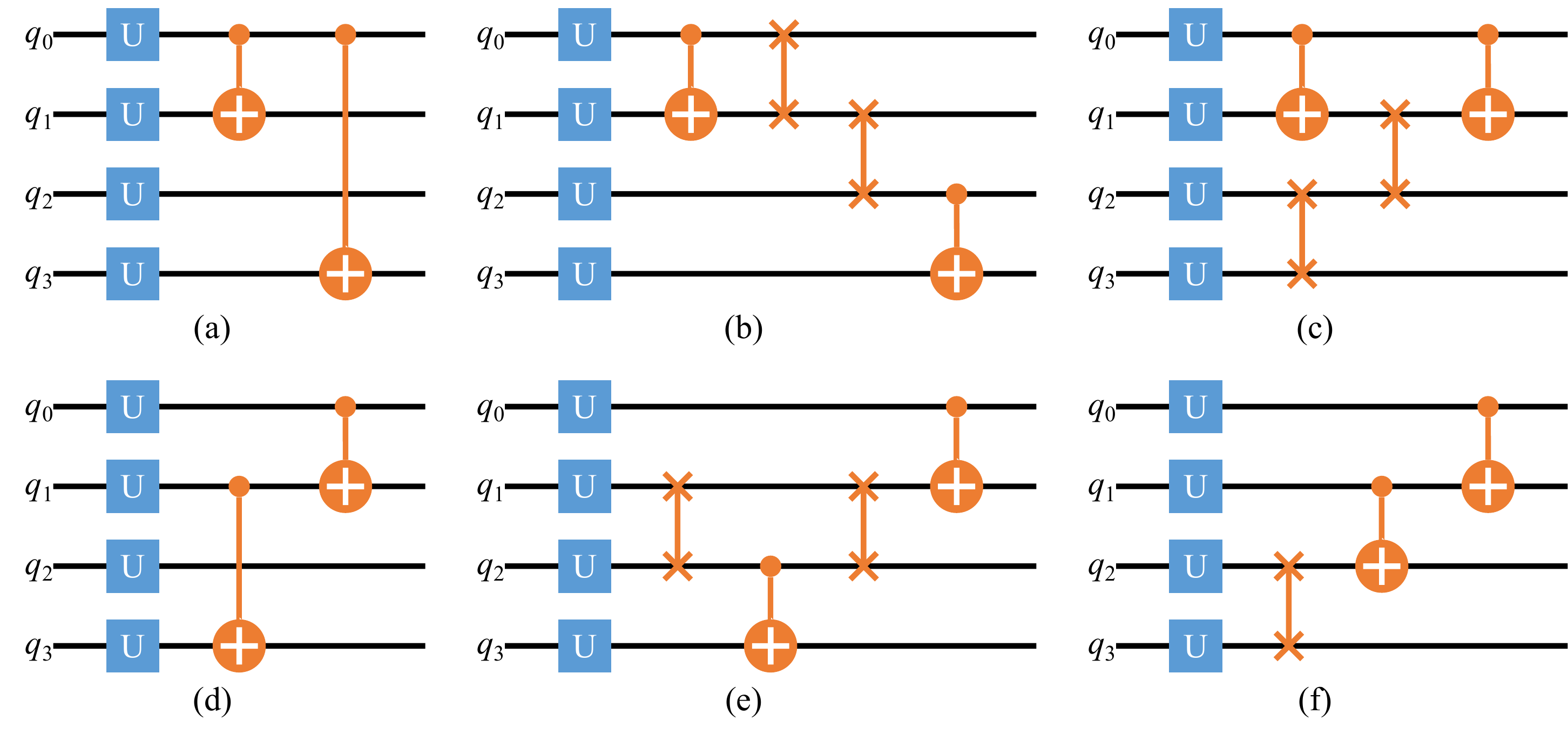}
    \caption{\label{fig:SWAPGateInsertionScheme}Comparison of the cost between different ways of SWAP gates insertion. (a) and (d) are two original quantum circuits. (b) and (e) show inserting swap gates in a inferior way, and the swap gates insertion method shown in (c) and (f) is advantage. Obviously, the quantum circuits shown in (c) and (f) are faster than the ones shown in (b) and (e) respectively.}
\end{figure*}

In general, the two target qubits of a two-qubit gate are not directly coupled, in order to perform the two-qubit gate operation, it is necessary to insert auxilliary SWAP gates. When a SWAP gate is inserted, an extra time will be added into the execution time of the quantum circuit. In addition, because there is a dependency between the two qubits used by the SWAP gate, even if one of the qubits is in an idle state, it cannot continue to execute the subsequent quantum gates. It must wait until the two qubits complete the SWAP gate together before continuing to execute the subsequent quantum gates. For example, to map quantum circuit as shown in Fig.\ref{fig:SWAPGateInsertionScheme}(a) into a quantum chip with linear coupling relationship (each qubit is directly coupled only to the two nearest neighbors, except the first and the last one, see Fig.\ref{fig:QuantumChipTopology}). To execute the CNOT gate applied on $q_0$ and $q_3$, it is necessary to insert SWAP gates to make $q_0$ and $q_3$ coupled. If we insert SWAP gates on ($q_0$, $q_1$) and ($q_1$, $q_2$)(b), problem of the coupling relationship will be solved. However, it also makes execution time longer than before. Because of the dependence between quantum gates, inserting a SWAP gate may cause longer execution time. But if we insert SWAP gates on ($q_2$, $q_3$) and ($q_1$, $q_2$) as shown in Fig.\ref{fig:SWAPGateInsertionScheme}(c), the quantum circuit will run faster than the quantum circuit shown in Fig.\ref{fig:SWAPGateInsertionScheme}(b). Since there is no dependence between the SWAP gate applied on ($q_2$, $q_3$) and the CNOT gate applied on ($q_0$, $q_1$), they can be executed simultaneously. Execution time of the quantum circuit shown in Fig.\ref{fig:SWAPGateInsertionScheme}(b) is $t_\text{U}+2t_{\text{SWAP}}+2t_{\text{CNOT}}$, and the execution time of the quantum circuit shown in Fig.\ref{fig:SWAPGateInsertionScheme}(c) is $t_\text{U} + 2t_{\text{SWAP}} + t_{\text{CNOT}}$. It is obvious that the quantum circuit shown in Fig.\ref{fig:SWAPGateInsertionScheme}(c) is faster about the time of a CNOT gate than the one shown in Fig.\ref{fig:SWAPGateInsertionScheme}(b).

 Not only that, the insertion of SWAP gates in quantum circuits may affect the coupling relationship of subsequent two-qubit gates, because the coupling relationship will change when we insert SWAP gates, so the CNOT gate that acted on a pair of coupled qubits may be not work. For example, mapping the quantum circuit in Fig.\ref{fig:SWAPGateInsertionScheme}(d) into a quantum chip with linear architectural by adding a SWAP gate to ($q_1, q_2$) as Fig.\ref{fig:SWAPGateInsertionScheme}(e) makes the next CNOT gate act on ($q_0, q_2$) so that the CNOT gate cannot be executed. It is necessary to employ an excess SWAP gate added to ($q_1, q_2$). However, if the SWAP gate is inserted on ($q_2, q_3$), it will only need one SWAP gate, such the quantum circuit shown in Fig.\ref{fig:SWAPGateInsertionScheme}(f). In this way, the quantum circuit will faster the time of a SWAP gate than the one shown in Fig.\ref{fig:SWAPGateInsertionScheme}(e).

\section{\label{sec:ArchitectureOfBQA}ARCHITECTURE OF BQA}

In this section, we go through the process to run a quantum program on a quantum computer. First, compiling a quantum program developed with high level programming languages into corresponding quantum circuit. Then reconfigure the quantum circuit with gates supported by the target quantum chip. Next, we should optimize the quantum circuit that consists of basic-gate set of the quantum chip according to the topology of the quantum chip to make it run well on the quantum chip. Last, the quantum circuit after optimizing is compiled into microwave pulses that are used to control physical-qubits on the quantum chip. In this section, we introduce the swap insertion architectural which is in the third step of running a quantum program.

% \subsection{Architecture of Optimization}
\begin{figure*}[htp]
    \centering
    \includegraphics[width=0.90\textwidth]{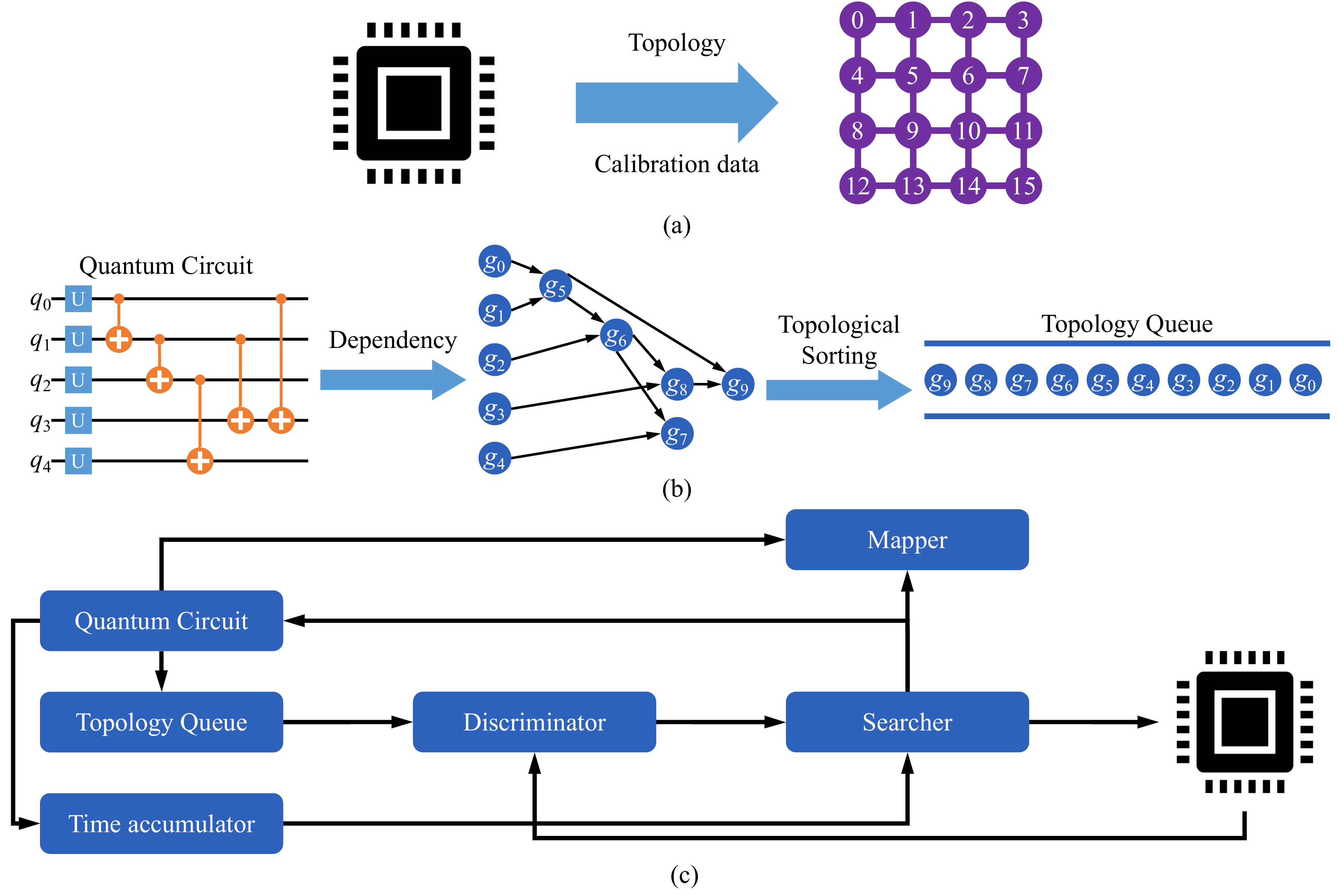}
    \caption{\label{fig:ArchitectureOfOptimization}(a) is the graph model of the quantum chip using the topology of the quantum chip and the calibration data. Each node in the graph represents a qubit on the chip, and each edge represents the two qubits connected by it are directly coupled. (b) is a directed acyclic graph (DAG) constructed by the dependencies between quantum gates, where the quantum gate represented by each node depends on the quantum gate represented by its precursor node. To keep the dependence between quantum gates, it is necessary to do topology sorting on the DAG. (c)The architecture of BQA consists of topology queue, discriminator, searcher, mapping table and time accumulator five parts. The topology queue makes sure the reliability of the dependence between quantum gates. The discriminator determines whether the quantum gate needs to be inserted into swap gates. The searcher finds the best way to insert swap gates. The time accumulator is employed to count the real time of qubits. The mapping table is used to correct the qubits position of qubit after inserting swap gates.}
\end{figure*}

Firstly, we need to establish models of qubits and quantum circuit. As shown in Fig.\ref{fig:ArchitectureOfOptimization}(a), we receive topology and calibration data of qubits from  the quantum chip and establish the abstract graph of the quantum chip. In the graph, each node represents a qubit and each edge means that the two qubits connected with this edge are directly coupled. To ensure the correctness of the quantum circuit, we can employ Directed Acyclic Graph (DAG) to make sure the dependence between quantum gates. In the DAG shown in Figure (5), each node represents a quantum gate, and it depends on the quantum gate represented by its predecessor nodes. In the DAG shown in Fig.\ref{fig:ArchitectureOfOptimization}(b), each node represents a quantum gate and depends the quantum gates  indicated by its precursor nodes. After constructing the DAG, we make a topological sorting to the DAG (choose a node without precursor node and push it into the queue, then remove all edges start from it until there is no node in the DAG). After completing the above preparations, we can optimize quantum circuits according to the dependence between quantum gates.

As Fig.\ref{fig:ArchitectureOfOptimization}(c) shows, the architecture of optimization consists of four parts, \emph{Discriminator}, \emph{Searcher}, \emph{Mapper} and \emph{Time accumulator}. Nodes enter the discriminator from the queue, then for these nodes corresponding to single-qubit gate or two-qubit gate that are acting on directly coupled qubits, they are sent to time accumulator to calculate the time of qubits after this gate. Otherwise, if a gate is a two-qubit gate acting on two qubits that are not directly coupled, it will be sent to searcher by discriminator. The searcher searches the best way to insert a SWAP gate for it from the topology graph of quantum chips. Searcher updates information of the best swap gate searched by itself to the quantum circuit and mapper. Gates on the qubits which are inserted swap gate are changed the qubits where they are allocated. Meanwhile, the information of gates the information of the quantum gates is synchronized to the nodes that represent them. Then the quantum circuit updates the time in time accumulator of the qubits inserted into the swap gate just inserted. Repeat these until there is no node in the queue. Finally, the output is fed to the mapper, and it is corrected according to the mapping relationships in the mapper.

\subsection{Searcher}
The searcher is used to search a way to insert swap gates for two-qubit gates which are not on directly coupled qubits. When the node at the head of the queue presents a two-qubit gate not acting on directly coupled qubits, it will not outbound team and the information about the gate which includes the control qubit and target qubit will be sent to the searcher. Firstly, searcher finds first steps and last steps of all paths from control qubit to target qubit from the graph of the topology of the quantum chip. Then the Searcher calculates the costs of inserting a swap gate for all steps gotten from topology graph by the cost function present in last subsection and inserts a swap gate according to the smallest one. The qubit information of nodes in queue will change when the swap gate is inserted into the quantum circuit. If the node at the head of the queue is on coupled qubits, it will be out of the queue and sent into the time accumulator.

\subsection{\label{subsec:TimeAccumulator}Time Accumulator}
The time accumulator is employed to denote the time of each qubit in real time. When a gate can work directly, the time accumulator will add its time to corresponding qubits. The time accumulator can help searcher appraise the business of qubits significantly. The way to count time of qubits must be divided into two situations because for single-qubit gate and two-qubits gate, the way to accumulate time is different. There is no dependence between for single-qubit gates, however, two-qubits gate are not. Therefore, the time of single-qubit gates can be add to the corresponding qubit directly, as shown in Eq.\eqref{con:acc_u},
\begin{equation}
    t_{q_{i},n+1} = t_{q_{i},n} + t_{q_{i},u}, \qquad (n=0, 1, \cdots)
    \label{con:acc_u}
\end{equation}
where  $q_i$ is the $i$-th qubit in the quantum circuit, $t_{q_{i},n}$ is the $i$-th qubit’s time after adding the $n$-th gate, $t_{q_{i},u}$ is the time of quantum gate $u$ on $q_i$. For two-qubit gates, there is dependence between two qubits, so the time of the two-qubit gate must be accumulated with the larger one of the time of two corresponding qubits, as shown in Eq.\eqref{con:acc_cnot},
\begin{equation}
    \begin{cases}
        t_{q_{i}, n+1} = \max (t_{q_{i}, n}, t_{q_{j}, m}) + t_{(q_{i}, q_{j}), CNOT}\\
        t_{q_{i}, m+1} = \max (t_{q_{i}, n}, t_{q_{j}, m}) + t_{(q_{i}, q_{j}), CNOT}
    \end{cases}
    \label{con:acc_cnot}
\end{equation}
where $m, n=0,1,2, \cdots$ and $t_{(q_{i}, q_{j}), CNOT}$ is the time of CNOT gate on qubits $i$ and $j$.

\subsection{Mapping Table}
By searcher’s work, the quantum circuit can work on the quantum chip, however, the result can not output directly. Because many swap gates are inserted, when we map the quantum circuit into the quantum chip. The swap gates solve the coupling problem between qubits, but they also lead to misalignment of qubits’ positions. Therefore, the output quantum state is not the one needed by us. Traditionally, this problem is solved by inserting swap to make the qubits locate on their start positions. The output state can be corrected truly in this way, but some overhead swap gates needed to be inserted and it makes the effect of optimization of BQA reduced. Therefore, mapping table is employed to correct the qubits’ position immediately.

\section{\label{sec:HeuristicSearch}METHOD OF HEURISTICALLY SEARCH}
Through the introduction of the motivation of swap gate insertion in subsection \ref{subsec:ChallengesOfSWAPInserting}, we know that when a swap gate is inserted, working time of the qubits acted on by the inserted SWAP gate will increase $t_{\text{SWAP}}$ immediately and the CNOT gates after it may need more swap gates to work. Therefore, there are two key issues to consider during inserting SWAP gate: the cost of the swap gate and overhead swap gates caused by it. To optimize swap gate insertion, it is necessary to overcome the two key issues. The swap gate depends on the evolution time of the qubit which is determined by the hardware. It can hardly be optimized by software. However, quantum circuits have extremely high parallelism, it means that quantum gates can be executed if there is no dependence. It is an opportunity that we can take the advantage of the high parallelism of quantum circuits to hide the time of swap gates or other gates. For example, in the quantum circuit shown in Fig.\ref{fig:SWAPGateInsertionScheme}(a), while the first CNOT gate on ($q_0, q_1$) is working, $q_2$ and $q_3$ are free. Inserting a swap gate on ($q_2, q_3$) as shown in Fig.\ref{fig:SWAPGateInsertionScheme}(c) the first CNOT gate on ($q_0, q_1$) is hidde by the inserted swap gate. We call the cost caused by swap gate itself as front cost.

The second key issue is the newly inserted SWAP gate may break the coupling relationship of some two-qubit gates that are originally acting on a pair of directly coupled qubits. To avoid this issue, we can directly count the overhead swap gates needed to be inserted when a swap gate is inserted. For example, for the quantum circuit shown in Fig.\ref{fig:SWAPGateInsertionScheme}(d), if we insert a swap gate on $q_1, q_2$, the second CNOT gate will not allocate on a pair of directly coupled qubits, it is necessary to insert another SWAP gate before performing the second CNOT gate, see Fig.\ref{fig:SWAPGateInsertionScheme}(e) However, if the swap gate is inserted on $q_2, q_3$, there will no swap gate needed to be inserted after it. We call this key issue as backend cost.

\begin{figure}[htp]
    \centering
    \includegraphics[width=.4\textwidth]{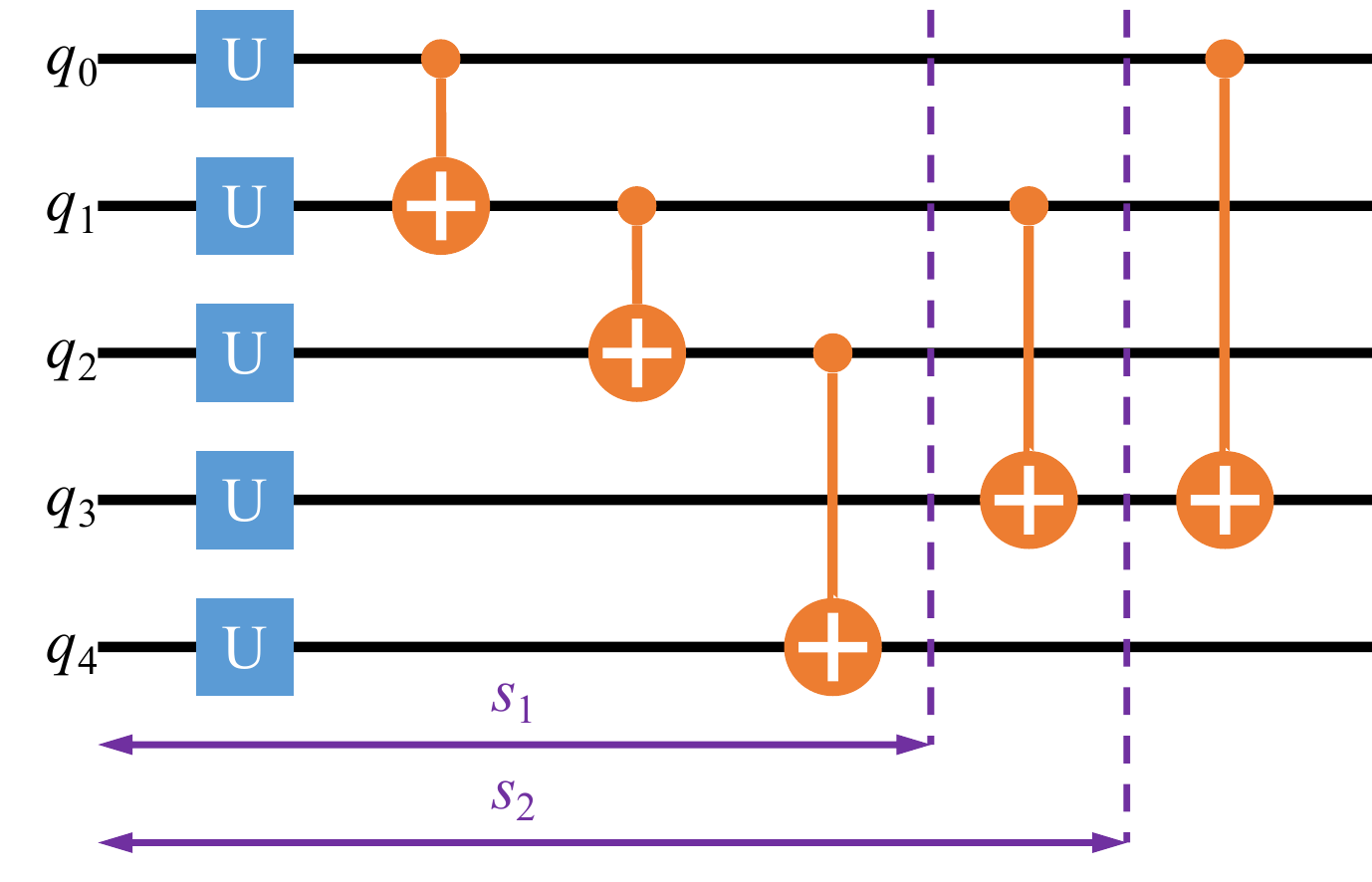}
    \caption{\label{fig:Subcircuit}The quantum circuit on a linear topology quantum chip is divided by the CNOT gates which are not on coupled qubits. In this figure, $s_1, s_2$ are subcircuits of the quantum circuit.}
\end{figure}

Based on the above characteristics, we design a heuristic search method.  In a quantum circuit, every qubit without two-qubit gate is parallel. They can be seen as processes in a parallel system and two-qubit gates are seen as the dependence between processes. For a parallel system, if there is no dependence between processes, the processes will have no free time. However, it is difficult for a parallel to avoid all dependence. Although the free time caused by dependence is waste of resources, in quantum circuits it can be used to insert swap gates. In a parallel system, the running time of the system is determined by the longest running process. For quantum circuits, the running time of the quantum circuit is the running time of the longest running qubit. Due to swap gates are only needed by CNOT gates not on coupled qubits, we should only consider about them. As shown in Fig.\ref{fig:Subcircuit}, a quantum circuit can be divided into several subcircuits by CNOT gates which are not on coupled qubits. After inserting swap gates, the time of subcircuit which can be calculated as shown in Eq.\eqref{con:front_cost} can present cost of the swap gates inserted just.
\begin{equation}
    Cost_{front} = \max \limits_{i}(t_{q_i,s_j})
    \label{con:front_cost}
\end{equation}
where $t_{q_i,s_j}$ is the time of the $i-th$ qubit in the $j-th$ subcircuit. Lower value of the front cost means hiding time more successful. To solve the backend cost, it is necessary to decrease the number of overhead swap gates caused by inserting swap gates. Therefore, we should not only consider the immediate CNOT gate, but also the future CONT gates. A counter which is used to count the number of swap gates needed by future CNOT gates should be employed to choose the swap gate which causes least backend cost. The backend cost can be calculated as follow:
\begin{equation}
    Cost_{backend} = \sum_{i}^{n} \text{SWAPNUM}(\text{CNOT}_i)
    \label{con:backend_cost}
\end{equation}
where $\text{SWAPNUM}(\text{CNOT})$ is the number of swap gates needed by the CONT gate, $n$ is the number of CNOT gates out of the subcircuit. The cost of inserting a swap gate is determined by both its front cost and backend cost. So, the cost of inserting a swap gate can be presented by the sum of the front cost and backend cost. However, when the subcircuit is small, the front cost may be small, and the backend may be large. With the expansion of the subcircuit, the front cost will be larger, and the backend will be smaller. It hardly has a time that front cost and backend cost are in the same order of magnitude. It is necessary to normalize to front cost and backend cost. The cost is only used to choose the best way to insert swap gate, its absolute value is not important. Therefore, we take the ratio of the qubit value with the largest time to the average of the times of all qubits as front cost. And the mean value of the number of swap gates needed by the CNOT gates out of the subcircuit is used to present the backend cost. The heuristic function is shown as Eq.\eqref{con:cost}.
\begin{align}
    cost & = \frac{\max\limits_{i}(t_{q_i,s_j})}{mean(t_{q_i,s_j})}  + \frac{\sum_{i}^{n} SWAPNUM(CNOT_i)}{n}
    \label{con:cost}
\end{align}

\section{\label{sec:Evaluation}EVALUATION}
\begin{table}
	\centering
	\caption{Evaluation time of CNOT gates on qubits}
	\label{tab:evaluationtime}
	\begin{tabular}{c c c}
		\hline \hline
		controlqubit &targetqubit &CNOT time ($\mu s$) \\
		$q_0$ &$q_1$ &0.540  \\
		$q_1$ &$q_2$ &0.739  \\
		$q_2$ &$q_3$ &0.675  \\
		$q_3$ &$q_4$ &0.512  \\
		$q_4$ &$q_5$ &0.540  \\
		$q_5$ &$q_6$ &0.540  \\
		$q_6$ &$q_7$ &0.64   \\
		$q_7$ &$q_8$ &0.65   \\
		$q_8$ &$q_9$ &0.248  \\
		\hline \hline
	\end{tabular}
\end{table}

\begin{figure}[htp]
    \centering
    \includegraphics[width=0.95\columnwidth]{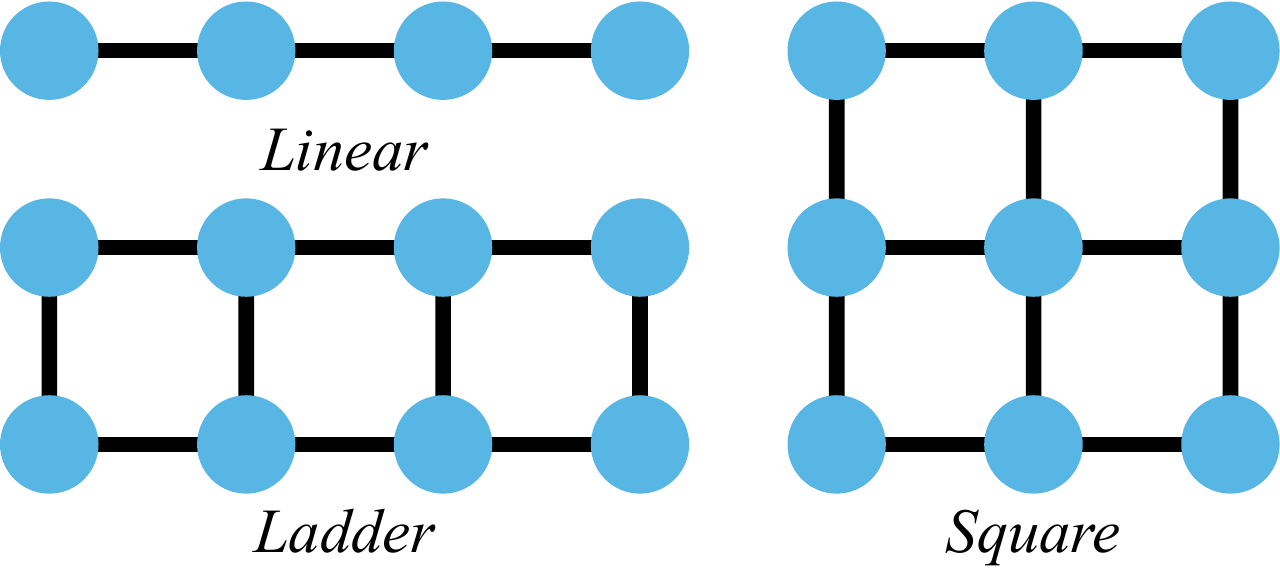}
    \caption{\label{fig:QuantumChipTopology}The three kinds of environment of the experiments.In the linear, qubits are connected one by one form the first one to the last one. The ladder can be seen as folding the linear in the middle so that each qubit relates to the qubit on or under it. And in the square, every qubit relates to the qubits around it. }
\end{figure}
In this section, we designed experiments to inspect the performance of the BQA. The BQA was implemented by Networkx and NumPy which are two libraries of python. The speed of quantum computer is so fast that we cannot measure the time of a quantum circuit accurately. What’s more, we only need to appraise the performance of the BQA. Therefore, it is not necessary to employ a quantum computer to execute the quantum circuits optimized by BQA. We use Python to construct a simulator based on numerical statistics to calculate the time of circuits. The simulator calculates the time of quantum circuits by accumulating time of quantum gates to their corresponding qubits like the time accumulator presented in subsection \ref{subsec:TimeAccumulator}. For single-qubit gates, the time of them is added to corresponding qubits directly as Eq.\eqref{con:acc_u}. For two-qubit gates, their time cannot be added to corresponding qubits without thinking about dependence. The simulator adds the time of a two-qubit gate to the larger one of two qubits’ time and assigns the value to both two qubits as Eq.\eqref{con:acc_cnot}. And the simulator can simulate three kinds of topology which are linear, ladder and square of quantum chips as shown in Fig.\ref{fig:QuantumChipTopology}. The evolution time of the single-bit gate on each qubit is $0.1 \mu s$, but the evolution time of the CNOT gate is different on different qubits. Therefore, we take the linear topology as an example and give the evolution time of CNOT gates on different qubits as shown in the Table \ref{tab:evaluationtime}. To inspect the performance of BQA clearly, the 0-3 level optimization of \emph{qiskit} are employed as the baseline of the experiment.

\subsection{Quantum Algorithm Optimization Evaluation}
\begin{figure*}[htp]
    \centering
    \includegraphics[width=0.90\textwidth]{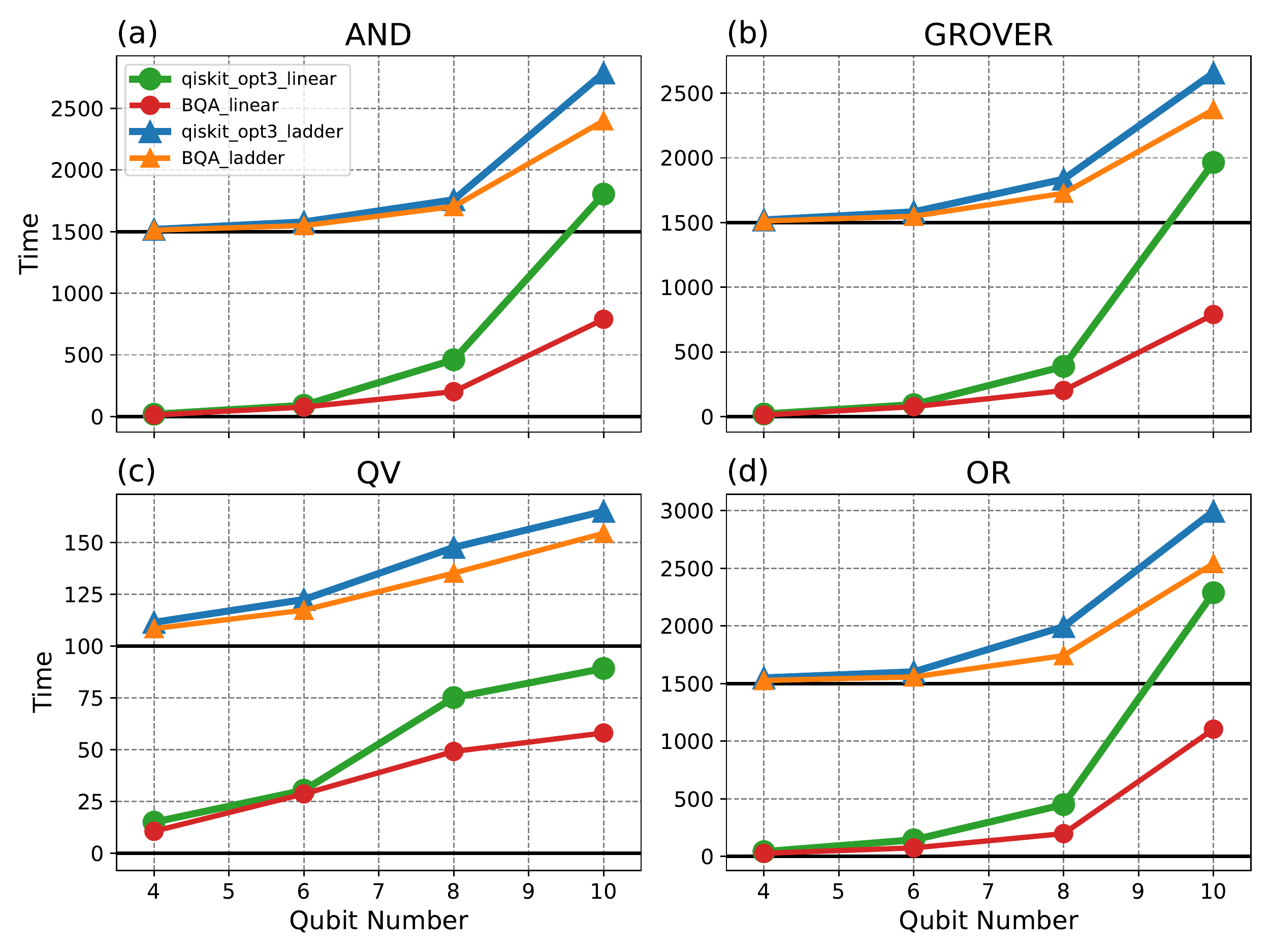}
    \caption{\label{fig:OptimizationResultsForSpecificAlgorithms} The execution time of several quantum algorithms versus the number of qubits. To avoid overlapping the curves, we shifted the blue and orange curves upward by a certain distance. The lower black horizontal solid lines correspond to the baselines of the red and green curves and the upper black horizontal solid lines correspond to the baselines of the orange and blue curves.}
\end{figure*}
To test the effect of BQA, we employed AND, OR, GROVER and Quantum Volume (QV) algorithms of 4, 6, 8, and 10 qubits as benchmarks. And the linear and ladder topology of quantum chips is used as environment. Fig.\ref{fig:OptimizationResultsForSpecificAlgorithms} shows the result of the experiment used to inspect the performance of BQA. As shown in Fig.\ref{fig:OptimizationResultsForSpecificAlgorithms}(a), \ref{fig:OptimizationResultsForSpecificAlgorithms}(b) and \ref{fig:OptimizationResultsForSpecificAlgorithms}(d), when the number of qubits of these algorithms is small, BQA is not better too much than \emph{qiskit}. However, with the number of qubits increasing, the distance between optimization effect of BQA and qiskit becomes larger and larger. As we can see, when the number of qubits is equal to ten, the performance of BQA is greatly better than the highest-level optimization of qiskit, especially on linear, the time of the circuit compiled by qiskit is about twice as much as the circuit compiled by BQA. For QV algorithm shown in Fig.\ref{fig:OptimizationResultsForSpecificAlgorithms}(c), the advantage of BQA is not as obvious as other algorithms, even though the number of qubits is ten, BQA is still greatly better than qiskit. Because the gate number of QV is small and increases slowly with the qubit number increasing. The gate number of QV does not have a large change when its qubit number changes. Therefore, with the number of qubits increasing, the optimization effect of BQA compared with qiskit only has little improvement. The influencing factors of the advantage of BQA also include the topology of the working environment for quantum circuit. From an overall perspective, algorithms on ladder needs less time than them on linear. And the distance between BQA and qiskit on ladder is smaller than linear. Because the couplers between qubits in ladder is more than linear so that fewer swap gates are needed to be inserted. It means that the need for optimization in the compilation of quantum circuits is less and optimization space is smaller. Therefore, BQA’s superiority compared with qiskit on ladder is less than linear. From these results, we can know that BQA makes a great effect on quantum circuit optimization and its performance is better than qiskit. And BQA has greater superiority on large scale quantum circuits and environment with less couplers between qubits.

\subsection{Random Quantum Gate Sequence Evaluation}
\begin{figure}[htp]
    \centering
    \includegraphics[width=0.95\columnwidth]{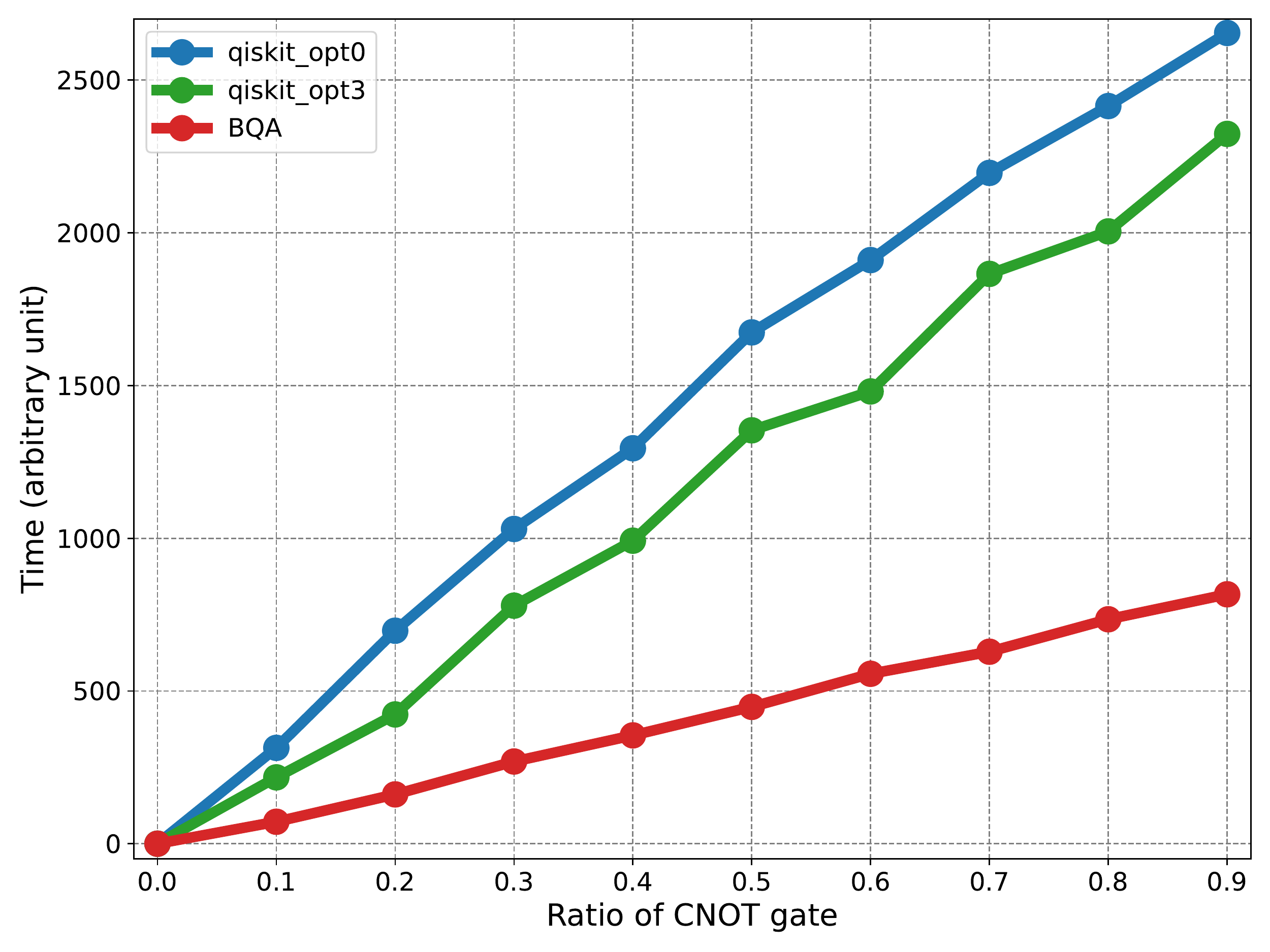}
    \caption{\label{fig:p_cx}Execution time versus the probabiltiy of CNOT gate. The qubit is 16 and the gate number is 950.}
\end{figure}

To analyze the factors affecting the performance of BQA clearly, random quantum gate sequence was employed by us. The random quantum gate sequence was generated by controlling qubit number, quantum gate num and the probability of CNOT gates. And it worked on the three kinds of topologies of the simulator respectively. The results of random quantum gate sequence is shown in Fig.\ref{fig:p_cx}.

\begin{figure}[htp]
    \centering
    \includegraphics[width=0.95\columnwidth]{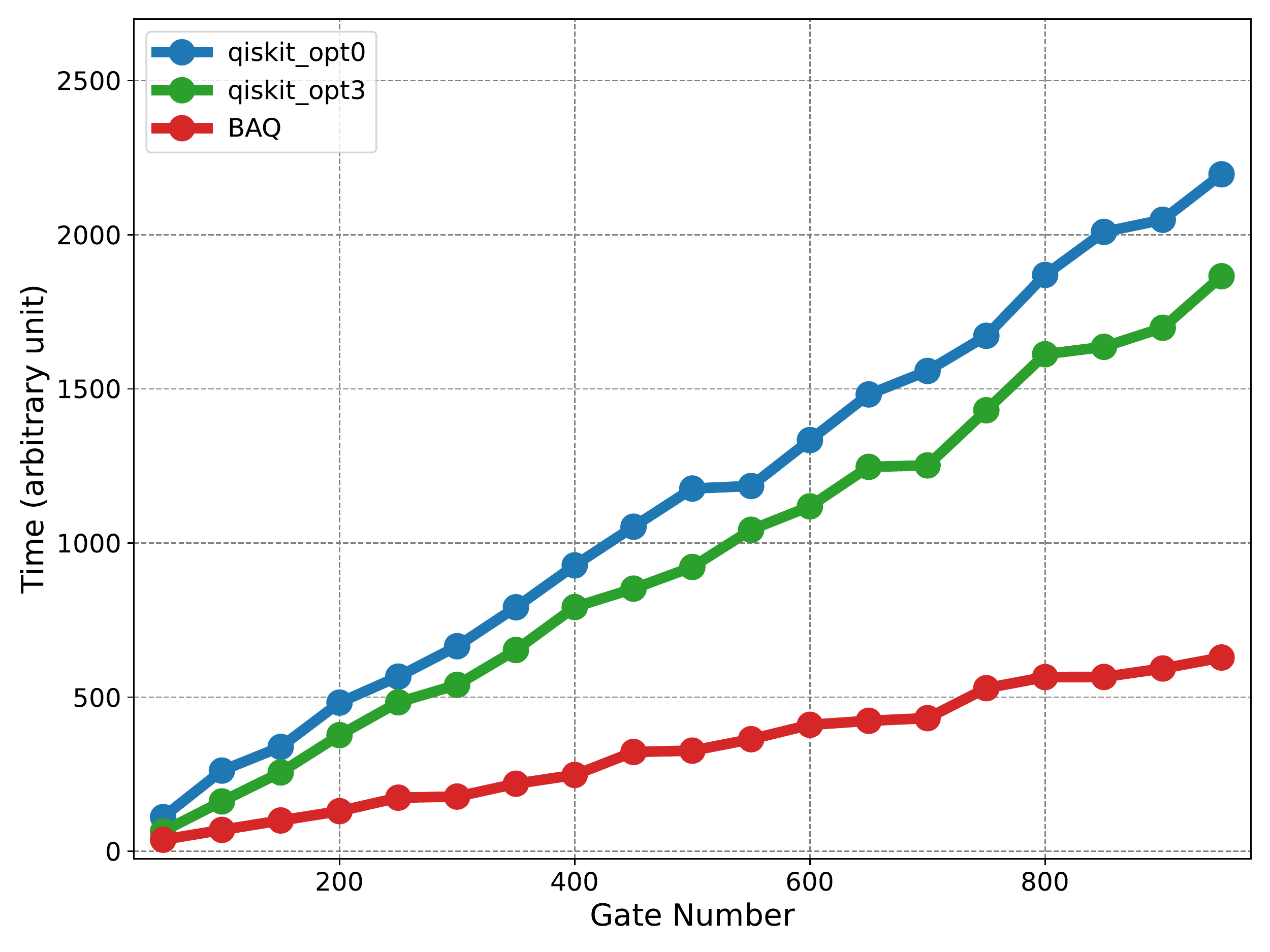}
    \caption{\label{fig:gate_num} Execution time versus the number of of quantum gates. }
\end{figure}

\begin{figure}[htbp]
	\centering
	\includegraphics[width=.95\columnwidth]{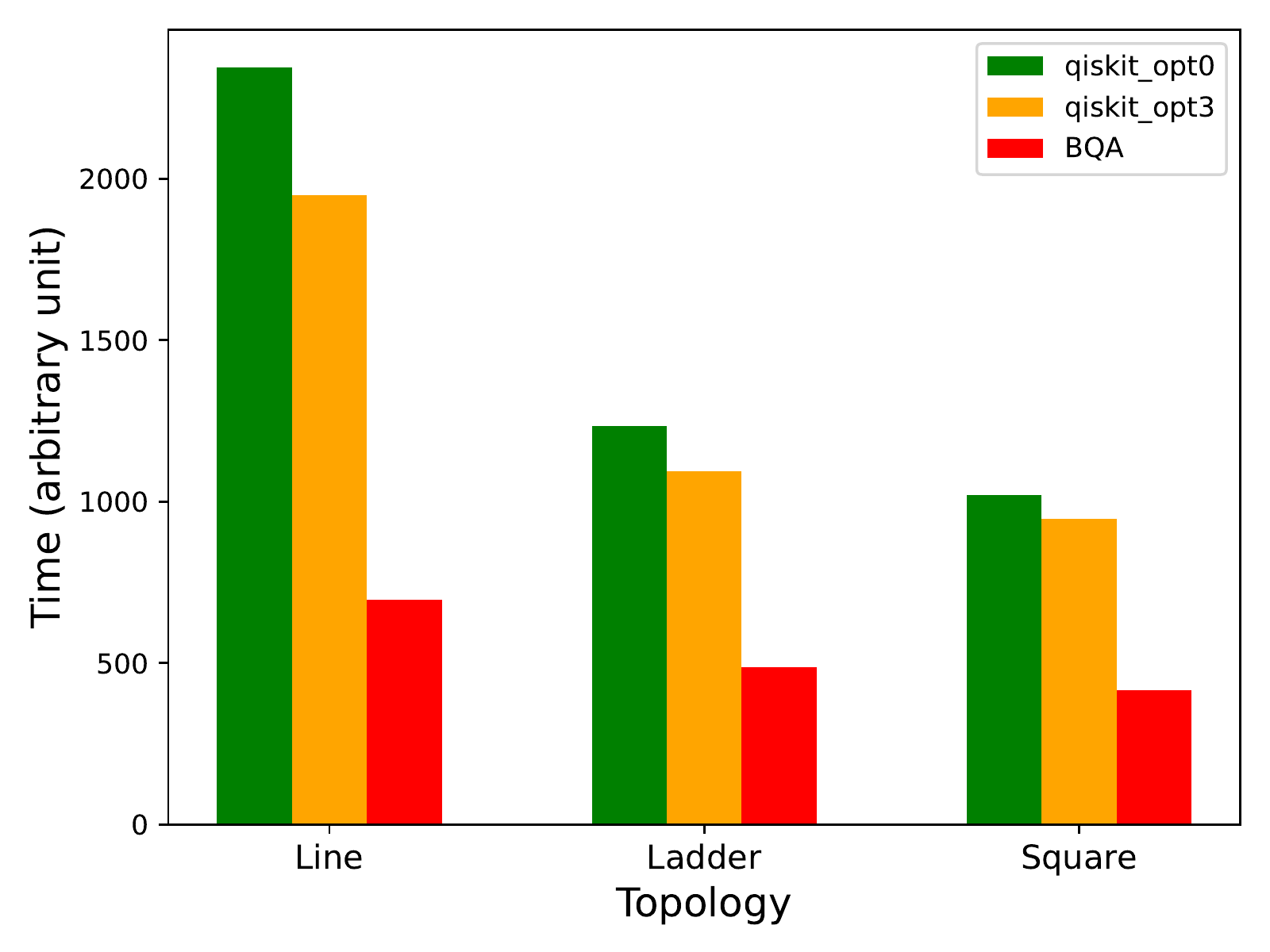}
	\caption{Execution time on three different topology quantum chips.}
	\label{fig:comparetopology}
\end{figure}

As we can see, BQA is still better than qiskit with the probability from 0.1 to 0.9. And with the probability of CNOT gates increasing, the distance between the performance of BQA and qiskit is larger and larger. When the probability of CNOT gate is 0.9, the time of the quantum circuit optimized by BQA is  only one thirds of the time of quantum circuit optimized by highest level of qiskit. Therefore, we can know that there are more two-qubit gates in quantum circuit the optimization effect of BQA is better. The time of random quantum gate sequences whose quantum gate number increases continue is shown as Fig.\ref{fig:gate_num}. In this figure, the time increases with quantum gate number increasing. And there are more quantum gates, the BQA more effective. According to the above two points, we drew a new conclusion that BQA has a greater effect on the quantum circuits having a large number of CNOT gates. As shown in Fig.\ref{fig:comparetopology}, we can obvious clearly that the time of the quantum circuits on square is shorter than ladder, the time of the quantum circuits on ladder is shorter than linear. However, the optimization effect of BQA on linear is better than ladder and on ladder is better than square. It can be seen that the optimization effect of BQA is enhanced as the coupling between qubits are reduced.
\section{\label{sec:Conclusion}CONCLUSION}
Quantum circuit optimization is more and more important for quantum computing. And swap gate insertion optimization is the core of quantum circuit optimization. It makes a significant influence to improve the performance of quantum computebut also the qubits’ business. Compared qiskit BQA make a great progress on quantum circuit optimization. BQA is more suitable for large-scale quantum circuits where a large number of SWAP gates need to be inserted. And for the topology effect with lower connectivity between qubits, BQA has a better optimization effect. We hope that our approach can provide a boost to the future development of quantum circuit optimization and quantum computing.

\nocite{*}
\bibliographystyle{elsarticle-num}
\bibliography{main.bib}

\end{document}